\documentclass[10pt,onecolumn,notitlepage]{revtex4-1}
\usepackage{amsfonts,amssymb}
\usepackage{graphicx}
\usepackage{color}
\usepackage{amsmath}
\usepackage{bm}
\usepackage[margin=1.1in]{geometry}

\newcommand{\pp}[2]{\frac{\partial#1}{\partial#2}}
\newcommand{\pppp}[2]{\frac{\partial^2#1}{\partial#2^2}}
\newcommand{\pppm}[3]{\frac{\partial^2#1}{\partial#2\partial#3}}

\begin{document}
\title{On the interpretation of the angular dependence \\of the FMR spectrum in heterogeneous ferromagnetic thin films}

\author{Maciej Kasperski and Henryk Puszkarski}
\affiliation{Surface Physics Division, Faculty of Physics,\\Adam Mickiewicz University\\
61-614 Pozna\'n, Umultowska 85, Poland}

\begin{abstract}
We demonstrate that a multi-peak FMR spectrum, with lines corresponding to resonance in different ferromagnetic regions of a heterogeneous thin-film sample, can collapse to a single-peak spectrum if there exists a particular field configuration, or the configuration of the external magnetic field with respect to the film surface, in which $dH_{\text{res}}/dM_{\text{eff}}=0$ within the region magnetically dominating in the sample.
\end{abstract}

\maketitle

\section{Introduction}
\label{sec:Introduction}

The ferromagnetic resonance (FMR) spectrum of thin films is known to evolve with the angular configuration of the applied magnetic field with respect to the film surface. In this evolution the multi-peak FMR spectrum often becomes a single-peak one in a certain angular configuration, to regain its multi-peak character beyond it. A careful analysis of this effect, in which a multi-peak FMR spectrum 'collapses' to a single-peak one, leads to the observation of two types of behavior of the collapsing spectrum. In one type the intensity of all the peaks except the first one diminishes progressively to vanish completely in the collapse configuration. Beyond this configuration angle the 'satellite' resonance lines emerge again in an unchanged order. The other type of collapse involves resonance positions rather than intensities: in the collapse configuration the lines shift to the position of the main peak, to reemerge in a reversed order beyond this configuration.
\par
The first type of collapse has been long known in the literature to be a surface effect due to the changes of the surface magnetic anisotropy (responsible for the surface spin pinning) with the configuration of the external field with respect to the surface of the thin film. In this pattern a complete collapse of the resonance spectrum occurs in a specific angular configuration (referred to as critical angle of the surface anisotropy) in which the surface anisotropy has no effect on the surface spin pinning, and the resonance precession of the 
all spins is homogeneous throughout the sample. In contrast, the other type of collapse is a   bulk effect that occurs as a result of the ferromagnetic resonance in separate regions of slightly different magnetic character in a heterogeneous thin-film sample. A heterogeneous magnetic structure is known to result in heterogeneous conditions of ferromagnetic resonance in the sample. Thus, in this interpretation each line in the multi-peak resonance spectrum corresponds to a resonance in a different region of the sample, and the collapse configuration of the external field is a particular configuration in which all the magnetically different regions participate in the resonance for resonance fields that differ only negligibly.
\par
The collapse of the FMR spectrum in this specific angular configuration of the external field in which the resonance heterogeneity of the sample is eliminated has not been thoroughly analyzed in the literature so far. The aim of this paper is to elucidate in detail the theoretical grounds of this effect.
\par
The paper is organized as follows. In Section II we recall the derivation of the universal Smit-Beljers-Suhl formula expressing the condition for ferromagnetic resonance to occur in a homogeneous bulk ferromagnetic sample. On the basis of this formula, in Section III we derive the configuration condition of resonance in thin films. In Section IV we use this condition for analyzing the configuration evolution of the resonance field in thin-film samples with a perpendicular uniaxial anisotropy. We formulate a universal condition for a multi-peak FMR spectrum to collapse into a single-peak one in such samples, and (in Section V) demonstrate the bulk character of this effect.

\section{Simplified derivation of the Smit-Beljers-Suhl resonance formula}\label{RfSBS}

\begin{figure}[p]
	\centering
		\includegraphics[width=4cm]{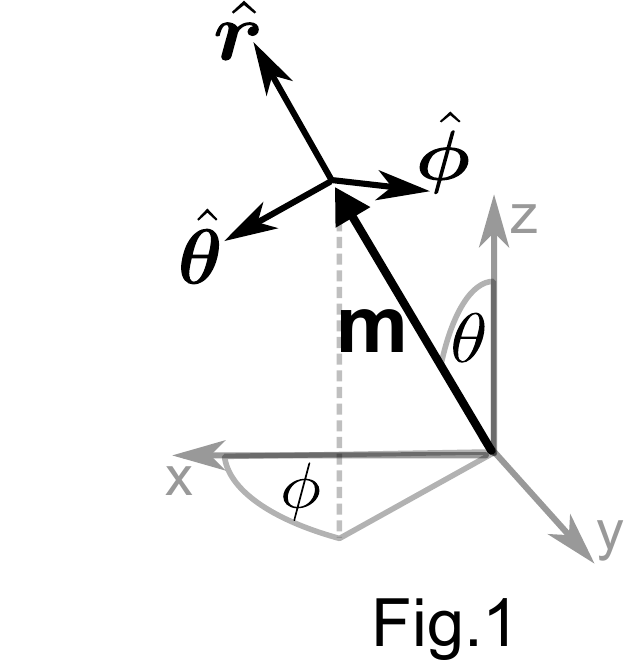}
	\caption{Local basis of spherical vectors $\hat{\bm{r}},\hat{\bm{\theta}},\hat{\bm{\phi}}$ related to vector $\bm{m}$.}
	\label{fig:Fig1}
\end{figure}

Let us consider the dynamics of a body with an angular momentum $\bm{J}$ and a magnetic moment $\bm{m}=\gamma\bm{L}$ collinear to it ($\gamma$ is the gyromagnetic ratio) in a magnetic field $\bm{H}$.
The field acts on the dipole to produce a torque $\bm{m}\times\bm{H}$ that sets the body in motion according to the equation:
\begin{equation}\label{.m}
	\dot{\bm{m}} = \gamma\bm{m}\times\bm{H}.
\end{equation}
Let us represent the vectors in the above equation in the local basis of orthonormal spherical vectors $\hat{\bm{r}},\hat{\bm{\theta}},\hat{\bm{\phi}}$ determined by the vector $\bm{m}(m,\theta,\phi)$ (obviously, $\bm{m}=m\hat{\bm{r}}$, see Fig.\ref{fig:Fig1} and comments in \cite{Gr}).
Thus:
\begin{equation}
	\dot{\bm{m}} = \gamma(m\hat{\bm{r}})\times(H_r\hat{\bm{r}} + H_{\theta}\hat{\bm{\theta}} + H_{\phi}\hat{\bm{\phi}}).
\end{equation}
Now, let us determine the components $H_{\theta}$ and $H_{\phi}$ of the magnetic field $\bm{H}$ by considering the change in the energy of the dipole with an infinitesimal rotation (we only consider rotation by the polar angle $\theta$).
The energy of a magnetic moment $\bm{m}$ in a magnetic field $\bm{H}$ is expressed by the equation:
\begin{equation}
	E(\bm{m}) = -\bm{m}\cdot\bm{H},
\end{equation}
which, represented in the basis $\hat{\bm{r}},\hat{\bm{\theta}},\hat{\bm{\phi}}$, becomes:

\begin{equation}
	E(\bm{m})  = -(m\hat{\bm{r}})\cdot(H_r\hat{\bm{r}} + H_{\theta}\hat{\bm{\theta}} + H_{\phi}\hat{\bm{\phi}}).
\end{equation}
As a result of the rotation of $\bm{m}$ by a small angle $\Delta\theta$ ($\bm{m} \stackrel{\Delta\theta}{\rightarrow} \bm{m}'=m\hat{\bm{r}}'$) the energy changes to:
\begin{equation}
\label{Em'}
	E(\bm{m'}) = -(m\hat{\bm{r}}')\cdot(H_r\hat{\bm{r}} + H_{\theta}\hat{\bm{\theta}} + H_{\phi}\hat{\bm{\phi}}).
\end{equation}
The vector $\hat{\bm{r}}'$ can be expressed as $\hat{\bm{r}} + \Delta\theta\hat{\bm{\theta}}$, so the above equation will become:
\begin{equation}
	E(\bm{m'}) = -m(\hat{\bm{r}} + \Delta\theta\hat{\bm{\theta}})\cdot(H_r\hat{\bm{r}} + H_{\theta}\hat{\bm{\theta}} + H_{\phi}\hat{\bm{\phi}}).
\end{equation}
The following relations are seen to occur:
\begin{equation}
	\Delta E = -mH_{\theta}\Delta\theta \quad\Rightarrow\quad H_{\theta} = -\left.\frac{1}{m}\pp{E}{\theta}\right|_{\theta,\phi},
\end{equation}
\begin{equation}
	 H_{\phi} = -\left.\frac{1}{m\sin\theta}\pp{E}{\phi}\right|_{\theta,\phi}.
\end{equation}
\par
Now, we can get back to the equation of motion~\eqref{.m} and write the cross product as the determinant:
\begin{equation}
  \left[
	 \begin{matrix}
	  \dot{m}\\
	  m\dot{\theta}\\
	  m\dot{\phi}\sin\theta\\
	 \end{matrix}
	\right]
	=
	\gamma
	\left|
	 \begin{matrix}
	  \hat{\bm{r}}&\hat{\bm{\theta}}&\hat{\bm{\phi}}\\
	  m&0&0\\
	  H_r&\left.\frac{-1}{m}\pp{E}{\theta}\right|_{\theta,\phi}&\left.\frac{-1}{m\sin\theta}\pp{E}{\phi}\right|_{\theta,\phi}\\
	 \end{matrix}
	\right| .
\end{equation}
Hence we obtain the system of equations:
\begin{equation}
\left\{
\begin{aligned}
 \dot{m} &= 0,\\
 \frac{m}{\gamma}\dot{\theta}\sin\theta &= \left.\pp{E}{\phi}\right|_{\theta,\phi},\\
 \frac{m}{\gamma}\dot{\phi}\sin\theta &= -\left.\pp{E}{\theta}\right|_{\theta,\phi}.
\end{aligned}
\right.
\end{equation}
This system of equations is to be solved in four steps. We (i)~determine the equilibrium angles~$\theta_0$ and~$\phi_0$ as the solution of the system of equations $\partial E/\partial\theta=\partial E/\partial\phi=0$; (ii)~expand $E$ and~$\sin\theta$ into a Taylor series at $\theta_0$ and~$\phi_0$, respectively; (iii)~assume that the time dependence of the angles has the form~$e^{i\omega t}$, and (iv)~exclude all the terms except the lowest-order ones (see \cite{Mo}). 
This procedure leads to the Smit-Beljers-Suhl formula \cite{SmBe,Su} that represents the ferromagnetic resonance condition:
\begin{equation}
 \frac{\omega}{\gamma} = \frac{1}{m\sin\theta_0}\sqrt{\left.\pppp{E}{\theta}\right|_{\text{eq}}\left.\pppp{E}{\phi}\right|_{\text{eq}} - \left(\left.\pppm{E}{\theta}{\phi}\right|_{\text{eq}}\right)^2}, \quad\gamma=\frac{g\mu_B}{\hbar},
\end{equation}
where the values of the derivatives correspond to the equilibrium angles~$\theta_0$ and~$\phi_0$.
(Note by the way that the radicand is the Hessian of the function~$E$ at the equilibrium point; its positive value is indicative of the existence of a minimum of the function at the point at which the first partial derivatives vanish.)
Since we shall henceforth consider samples with a magnetization~$\bm{M}$, the formulas derived above, with~$\bm{M}$ in place of~$\bm{m}$, will apply to the description of the magnetization dynamics in the studied case.

\section{Resonance condition in thin films with uniaxial anisotropy}\label{sec:Wrdcwzaj}

Let us consider a sample in which the free energy density $E$ is the sum of the Zeeman energy, the demagnetization energy and the uniaxial anisotropy energy:
\begin{equation}
\label{E}
	E = -\bm{M}\cdot\bm{H} + 2\pi(\bm{M}\cdot\bm{n})^2 - K(\bm{M}\cdot\bm{u}/M)^2,
\end{equation}
where $\bm{n}$ is a unit vector normal to the surface of the sample, and $\bm{u}$ is a unit vector oriented along the easy magnetization axis. The applied magnetic field and the magnetization of the sample will be henceforth represented as (see Fig.\ref{fig:Fig2}a):
\begin{figure}[p]
	\centering
		\includegraphics[width=10cm]{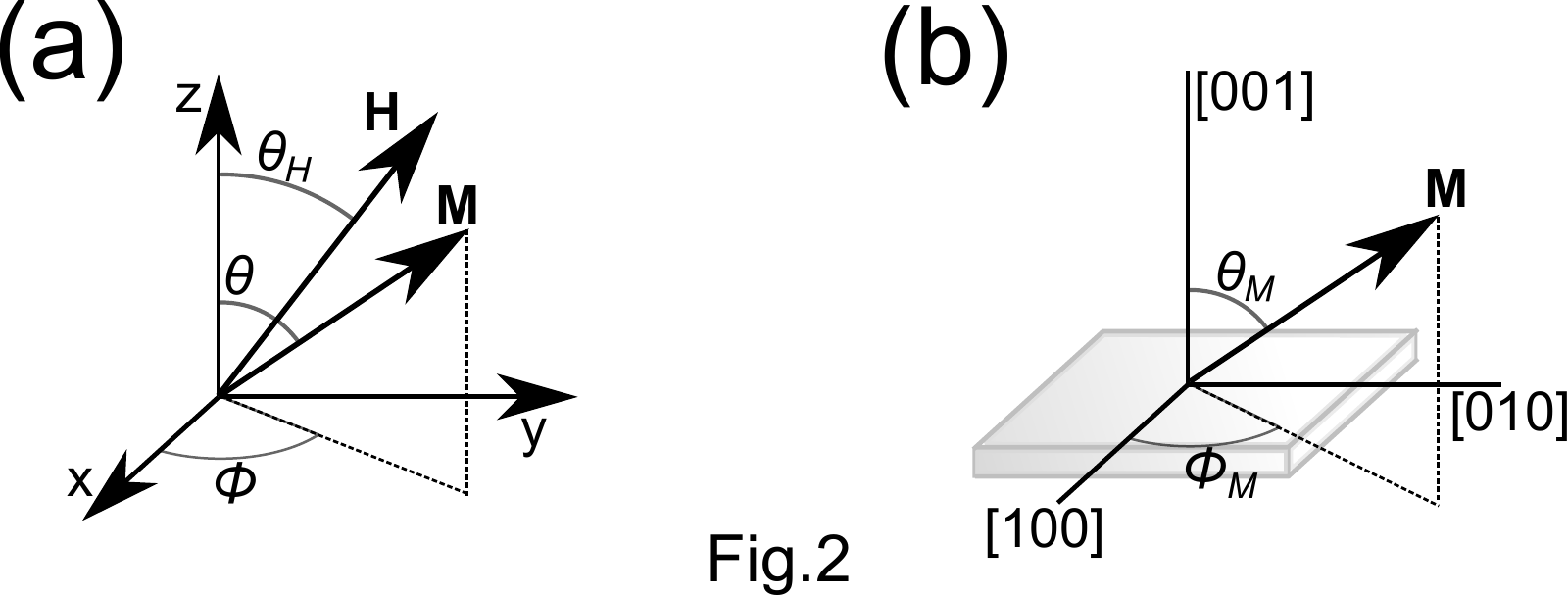}
 \caption{Definition of the angular configuration of the magnetization $\bm{M}$ and the applied magnetic field $\bm{H}$ with respect to the film surface.}
	\label{fig:Fig2}
\end{figure}

\begin{equation}
	\begin{aligned}
		\bm{M} &= M(\sin\theta\cos\phi, \sin\theta\sin\phi, \cos\theta)\\
		\bm{H} &= H(0, \sin\theta_H, \cos\theta_H).
	\end{aligned}
\end{equation}
We shall consider three special cases corresponding to three different choices of the easy axis ($[100],[010]$ or $[001]$, see Fig.\ref{fig:Fig2}b).
In each case we shall investigate the angular configuration dependence~$H_{\text{res}}(\theta_H)$ implied by the assumed form of the free energy density~\eqref{E}.

\subsection{Perpendicular uniaxial anisotropy, $\bm{u}=[001]$}
\label{sec:Uaip001}

In this case the unit vector~$\bm{n}$ normal to the surface of the sample and the unit vector~$\bm{u}$ oriented along the easy axis are identical:
\begin{equation}
	\bm{n} = \bm{u} = [001].
\end{equation}
Thus, the free energy density reads:
\begin{equation}
	E = -HM\left(\sin\phi\sin\theta\sin\theta_H+\cos\theta\cos\theta_H\right)\\
+(2\pi M^2 - K)\cos^2\theta.
\end{equation}
Let us determine the equilibrium conditions of the system. The first partial derivatives with respect to the angles $\theta$ and $\phi$ have the form:
\begin{equation}
	\left\{
	\begin{aligned}
		\pp{E}{\theta} &= -MH\left(\sin\phi\cos\theta\sin\theta_H-\sin\theta\cos\theta_H\right)-(2\pi M^2 - K)\sin2\theta\\
		\pp{E}{\phi} &= -MH\cos\phi\sin\theta\sin\theta_H.
	\end{aligned}
	\right.
\end{equation}
From the condition of vanishing of these derivatives we obtain the equations that determine the coordinates $\theta_M$ and $\phi_M$ of the equilibrium point (Fig.\ref{fig:Fig2}b):
\begin{subequations}\label{equi1}
	\begin{align}
		2H\sin(\theta_M - \theta_H) &= 4\pi M_{\text{eff}}\sin2\theta_M
		\label{equi1a}\\
		\cos\phi_M& = 0 \quad \Leftarrow \quad \phi_M = \pi/2
		\label{equi1b} ,
	\end{align}
\end{subequations}
where $M_{\text{eff}}$ is the effective magnetization, defined as:
\begin{equation}
	4\pi M_{\text{eff}} = 4\pi M - 2K/M .
\end{equation}
In this case the Smit-Beljers-Suhl formula becomes:
\begin{equation}\label{SBS1}
	\frac{\omega^2}{\gamma^2}=\left[H\cos(\theta_M-\theta_H)-4\pi M_{\text{eff}}\cos^2\theta_M\right]
\left[H\cos(\theta_M-\theta_H)-4\pi M_{\text{eff}}\cos2\theta_M\right].
\end{equation}
In order to plot the corresponding configuration dependence~$H_{\text{res}}(\theta_H)$ of the resonance field we must solve the system of two nonlinear equations~\eqref{equi1a} and~\eqref{SBS1}.

\subsection{In-plane uniaxial anisotropy, $\bm{u}=[010]$}
\label{sec:Uaip010}

In this case the unit vector~$\bm{n}$ normal to the surface and the unit vector~$\bm{u}$ oriented along the anisotropy direction are:
\begin{equation}
	\bm{n} = [001] \quad\text{and}\quad \bm{u} = [010].
\end{equation}
The free energy density is:
\begin{equation}
	E = -HM\left(\sin\phi\sin\theta\sin\theta_H+\cos\theta\cos\theta_H\right)+2\pi M^2\cos^2\theta - K\sin^2\theta\sin^2\phi.
\end{equation}
The first partial derivatives with respect to the angles~$\theta$ and~$\phi$ read:
\begin{equation}
	\left\{
	\begin{aligned}
		\pp{E}{\theta} =& -MH\left(\sin\phi\cos\theta\sin\theta_H-\sin\theta\cos\theta_H\right) \\ 
&- 4\pi M^2\sin\theta\cos\theta -2K\sin^2\phi\sin\theta\cos\theta\\
		\pp{E}{\phi} =& -MH\cos\phi\sin\theta\sin\theta_H - 2K\sin\phi\cos\phi\sin^2\theta .
	\end{aligned}
	\right.
\end{equation}
Thus, the equilibrium direction of magnetization is determined by the equations:
\begin{equation}\label{equi2}
	\left\{
\begin{aligned}
		0 & = 2H(\sin\phi_M\cos\theta_M\sin\theta_H-\sin\theta_M\cos\theta_H) + (4\pi M+\frac{2K}{M}\sin^2\phi_M)\sin2\theta_M \\
		0 & = \cos\phi_M\sin\theta_M\left(H\sin\theta_H+\frac{2K}{M}\sin\phi_M\sin\theta_M\right).
\end{aligned}
	\right.
\end{equation}
The Smit-Beljers-Suhl formula for this case is:
\begin{equation}
	\frac{\omega^2}{\gamma^2}=\left[H\cos(\theta_M-\theta_H)-\left(4\pi M+\frac{2K}{M}\right)\cos2\theta_M\right]
	\left[H\frac{\sin\theta_H}{\sin\theta_M}+\frac{2K}{M}\right],
\end{equation}
or, equivalently:
\begin{multline}
	\frac{\omega^2}{\gamma^2}=\left[H\cos(\theta_M-\theta_H)-\left(4\pi M+\frac{2K}{M}\right)\cos2\theta_M\right]\\\times
\left[H\cos(\theta_M-\theta_H)-\left(4\pi M + \frac{2K}{M}\right)\cos^2\theta_M+\frac{2K}{M}\right].
\end{multline}
Note that this condition does not involve the azimuth angle $\phi$, since the equilibrium condition \eqref{equi2} requires again that $\phi_M=\pi/2$.

\subsection{In-plane uniaxial anisotropy, $\bm{u}=[100]$}
\label{sec:Uaip100}

In this case the unit vector~$\bm{n}$ normal to the sample surface and the unit vector~$\bm{u}$ parallel to the anisotropy direction are:
\begin{equation}
	\bm{n} = [001] \quad\text{and}\quad \bm{u} = [100].
\end{equation}
The free energy density reads:
\begin{equation}
	E = -HM\left(\sin\phi\sin\theta\sin\theta_H+\cos\theta\cos\theta_H\right)+2\pi M^2\cos^2\theta - K\sin^2\theta\cos^2\phi.
\end{equation}
The first partial derivatives of the energy density with respect to the angles $\theta$ and $\phi$ have the form:
\begin{equation}
	\left\{
	\begin{aligned}
		\pp{E}{\theta} =& -HM(\sin\phi\cos\theta\sin\theta_H-\sin\theta\cos\theta_H) \\
&-4\pi M^2\sin\theta\cos\theta-2K\cos^2\phi\sin\theta\cos\theta\\
		\pp{E}{\phi} =& -HM\cos\phi\sin\theta\sin\theta_H + 2K\sin\phi\cos\phi\sin^2\theta .
	\end{aligned}
	\right.
\end{equation}
Thus, the equilibrium direction of magnetization is determined by the angles $\theta_M$ and $\phi_M$ that fulfill the equations:
\begin{equation}
	\left\{
	\begin{aligned}
		0 =& -HM(\sin\phi_M\cos\theta_M\sin\theta_H-\sin\theta_M\cos\theta_H) \\
&-4\pi M^2\sin\theta_M\cos\theta_M-2K\cos^2\phi_M\sin\theta_M\cos\theta_M\\
		0 =& -HM\cos\phi_M\sin\theta_M\sin\theta_H + 2K\sin\phi_M\cos\phi_M\sin^2\theta_M.
	\end{aligned}
	\right.
\end{equation}
The Smit-Beljers-Suhl formula for this case has the form:
\begin{multline}
	\frac{\omega^2}{\gamma^2}
	=
	\left[H(\sin\phi_M\sin\theta_M\sin\theta_H+\cos\theta_M\cos\theta_H)
	- \left(4\pi M + \frac{2K}{M}\cos^2\phi_M\right)\cos2\theta_M\right]\\
	\times
	\left[\frac{H\sin\phi_M\sin\theta_H}{\sin\theta_M}+\frac{2K}{M}\cos2\phi_M\right]
	-\left(\frac{H\cos\phi_M\cos\theta_M\sin\theta_H}{\sin\theta_M}-\frac{2K}{M}\sin2\phi_M\cos\theta_M\right)^2
\end{multline}
or, equivalently:
\begin{multline}
	\frac{\omega^2}{\gamma^2}
	=
	\left[H(\sin\phi_M\sin\theta_M\sin\theta_H+\cos\theta_M\cos\theta_H)-\left(4\pi M + \frac{2K}{M}\cos^2\phi_M\right)\cos2\theta_M\right]
	\\\times
	\left[H(\sin\phi_M\sin\theta_M\sin\theta_H+\cos\theta_M\cos\theta_H)-	\left(4\pi M + \frac{2K}{M}\cos^2\phi_M\right)\cos^2\theta_M + \frac{2K}{M}\cos2\phi_M\right]
	\\-\left(\frac{2K}{M}\right)^2\sin^2\phi_M\cos^2\phi_M\cos^2\theta_M.
\end{multline}

\section{The existence of a resonance intersection point in heterogeneous thin films}
\label{sec:TheExistenceOfAResonanceIntersectionPointInHeterogeneousThinFilms}

\begin{figure}[p]
	\includegraphics[width=10cm]{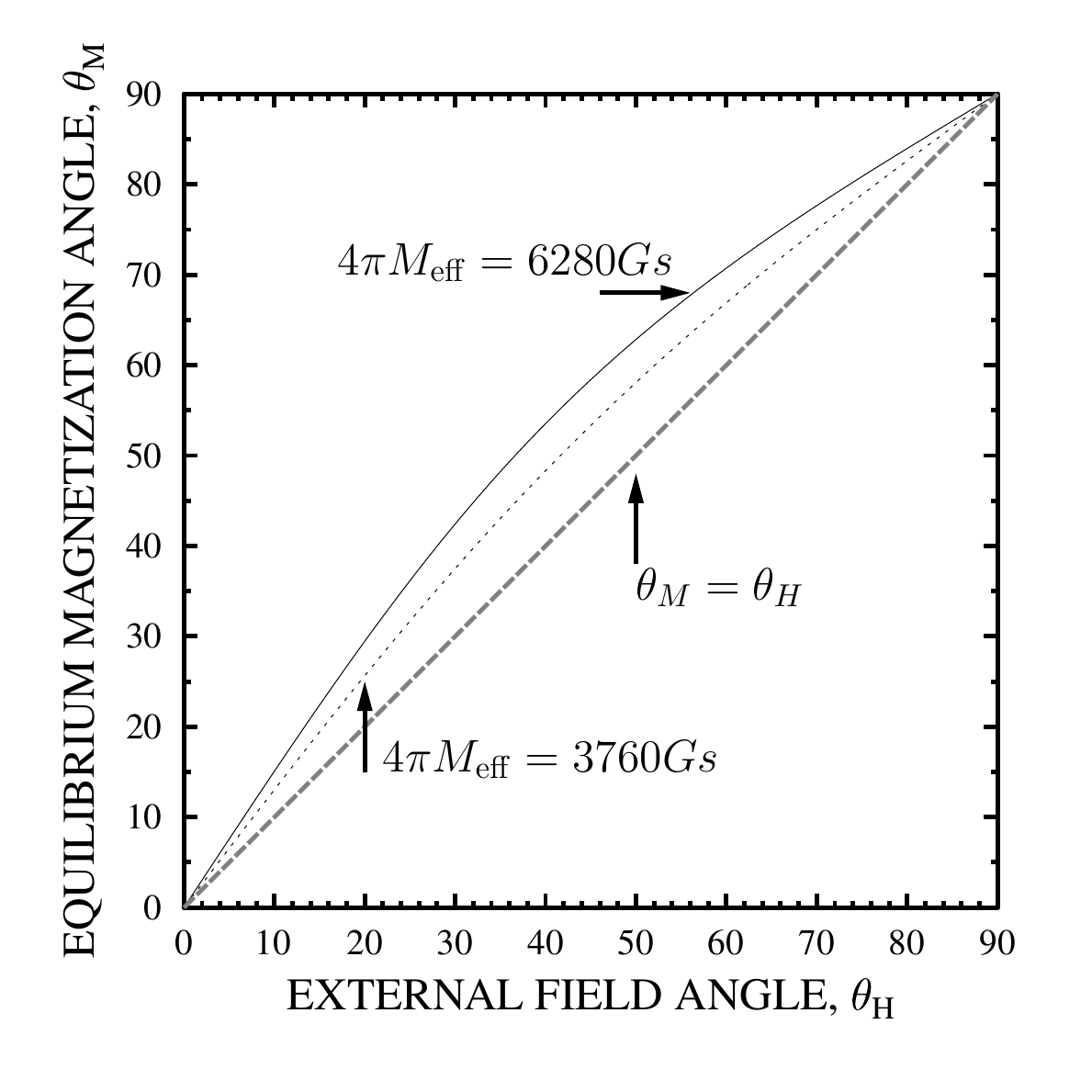}
	\caption{Equilibrium magnetization angle $\theta_M$  vs. external field angle $\theta_H$ determined for a thin film with perpendicular uniaxial anisotropy (f=33.5 GHz); two curves corresponding to two effective magnetization values are plotted for comparison.}
	\label{fig:FERROMAGNETIC_RESONANCE_A_Angle}
\end{figure}

In this Section we shall only consider the simplest of the three cases mentioned above, i.e. a thin film with perpendicular uniaxial anisotropy (case [001]); we leave the remaining two cases to be similarly discussed in a separate paper.
The condition \eqref{equi1a} allows to determine the equilibrium angle $\theta_M$ of the magnetization vector corresponding to a specific configuration of the external field $\theta_H$.
Figure \ref{fig:FERROMAGNETIC_RESONANCE_A_Angle} shows the consequent configuration dependence $\theta_M(\theta_H)$ for two values of $4\pi M_{\text{eff}}$.
The deviation of the magnetization angle $\theta_M$ from the field angle $\theta_H$ is seen to be the largest in the middle of the interval $[0,\pi/2]$ , and to grow with increasing value of $4\pi M_{\text{eff}}$; in this range the deviation is of ca. $10^{\circ}$ to $15^{\circ}$, far from being negligible!
\par
\begin{figure}[h]
	\centering
	\includegraphics[width=8cm]{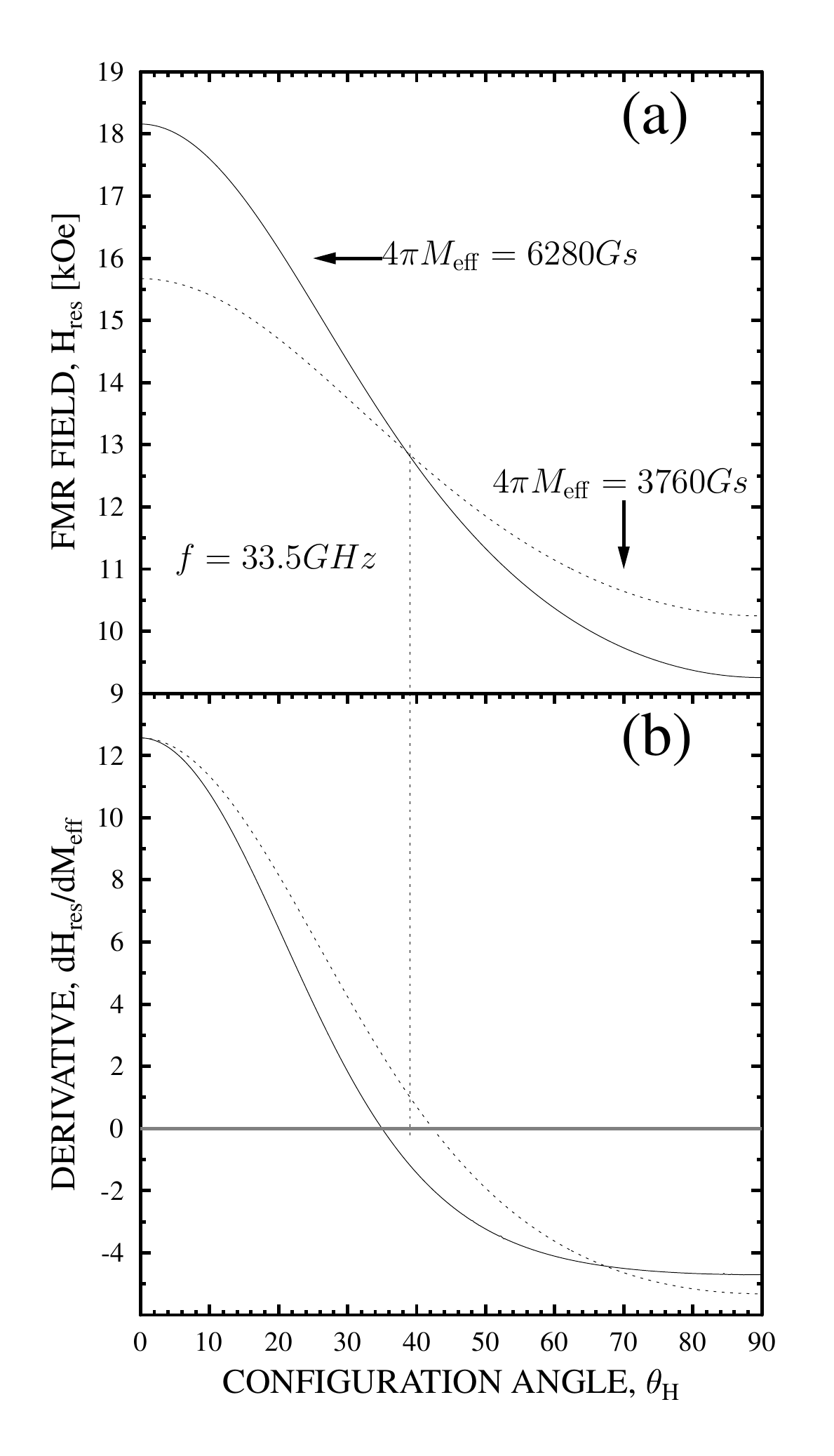}
	\caption{Configuration dependence of (a) the resonance field $H_{\text{res}}(\theta_H)$ and (b) its derivative $dH_{\text{res}}/dM_{\text{eff}}$ in a thin film with perpendicular uniaxial anisotropy.
(a)	Note the intersection of the resonance curves plotted for the two regions of different effective magnetization magnitude. The intersection point corresponds to the particular field configuration in which the sample produces a homogeneous resonance for the same field magnitude throughout its volume.
(b)	Note that the homogeneous resonance configuration lies in the range determined by the vanishing of the derivative $dH_{\text{res}}/dM_{\text{eff}}$ in each resonating region.}
	\label{fig:FERROMAGNETIC_RESONANCE_A_Field+DHdM}
\end{figure}

Now, let us assume that the thin film under consideration is heterogeneous and planarly stratified into regions with two different values of effective magnetization $4\pi M_{\text{eff}}$.
The resonance condition \eqref{SBS1} -- with the equilibrium condition \eqref{equi1a} used -- leads to two respective configuration resonance curves $H_{\text{res}}(\theta_H)$, see Fig.\ref{fig:FERROMAGNETIC_RESONANCE_A_Field+DHdM}a.
This means that for a configuration $\theta_H$ each of the ferromagnetic strata resonates for a different value of the external field.
This applies to the whole range of $\theta_H$ with the exception of one particular configuration corresponding to the point of intersection of the two curves.
In this particular configuration both regions -- in spite of their significantly different magnetic parameters! -- produce a resonance simultaneously for the same magnitude of the external field.
This implies homogeneous precession throughout the sample, as if its heterogeneity were completely eliminated in this configuration!
Thus, for this particular angle $\theta_H$ the sample can be regarded as \emph{dynamically homogeneous}.
The dynamic homogeneity (DH) is evidenced by the collapse of the two-peak resonance spectrum to a single resonance line.
Note that past this particular angle the two peaks reemerge, but the resonance sequence is reversed.
\par
The occurrence of a \emph{dynamic homogeneity angle} (DHA) is an intrinsic characteristic of the Smit-Beljers-Suhl (SBS) resonance formula.
Thus, we have grounds to ask whether the SBS equation can provide the basis for an analytical condition that would allow to predict the DHA angle for a ferromagnetic thin-film sample.
We think we have managed to formulate such a condition a posteriori on the basis of the numerical results shown in Fig.\ref{fig:FERROMAGNETIC_RESONANCE_A_Field+DHdM}b, presenting the configuration dependence of the derivative $\partial H/\partial M_{\text{eff}}$ determined for each of the ferromagnetic regions shown in Fig.\ref{fig:FERROMAGNETIC_RESONANCE_A_Field+DHdM}a.
The DHA angle is seen to be in the range delimited by the extreme configurations for which the considered derivative vanishes in each stratum.
The range shrinks with decreasing difference between the extreme values of $4\pi M_{\text{eff}}$ in the two regions.
Thus, we can expect that the DHA angle will be determined with a good approximation by the condition of vanishing of the derivative:
\begin{equation}\label{DHA}
	\pp{H_{\text{res}}(\theta_H)}{M_{\text{eff}}} = 0
\end{equation}
in the region the magnetic properties of which predominate in the sample.

\section{Conclusion}
\label{sec:Conclusion}

It is worthy of notice that the above-discussed collapse of the FMR spectrum in a particular field configuration defined by the DHA angle is a dynamic \textbf{volume} effect, since only bulk quantities figure in the adopted expression for free energy density, and the presence of the surface is only manifested in the shape anisotropy of the sample.
Thus, the energy of surface anisotropy was not taken into account in our considerations.
This is an important statement, since also the surface anisotropy by itself can cause a multipeak SWR spectrum to collapse to a single resonance line in a certain configuration, referred to as \emph{critical angle} \cite{Pu}, of the applied field with respect to the film surface.
However, there is a substantial difference between the collapse of an FMR spectrum and that of an SWR spectrum.
In the former the resonance line positions merge to form a single line at the DHA angle, while the collapse of an SWR spectrum consists in the suppression of the intensity of all the lines except the main one at the critical angle.
This is due to the different nature of the two effects: the essence of the FMR collapse is the attainment of homogeneous resonance dynamics throughout the \emph{volume} of the sample, whereas the SWR effect consists in the elimination, in the \emph{critical} field configuration, of the impact of the surface anisotropy on the spin dynamics in the thin film.
Therefore, we shall refer to the latter configuration as \emph{surface critical angle} (SCA).
The SCA is sensitive to the conditions on the surface of the sample and can vary with them, while the DHA, only determined by the bulk characteristics that enter the condition \eqref{DHA}, is a fixed parameter of the sample.
It should be emphasized that the above-mentioned differences between the two effects can be used in practice as a criterion that allows to determine whether a collapse of a multipeak FMR spectrum observed in a thin film is of surface or bulk character.

\section{Acknowledgments}
\label{sec:Acknowledgments}

This study was supported by the Polish Ministry of Science and Higher Education, Grant No. N N202 1945 33.


\end{document}